# Precision Spectroscopy of Polarized Molecules in an Ion Trap[*]

H. Loh[1†], K. C. Cossel[1], M. C. Grau[1], K.-K. Ni[1‡], E. R. Meyer[2], J. L. Bohn[1], J. Ye[1†], E. A. Cornell[1†]

[1]JILA, NIST and University of Colorado, and
Department of Physics, University of Colorado, Boulder, CO 80309-0440, USA.

[2]Department of Physics, Kansas State University, Manhattan, KS 66506-2601, USA.

[†]Correspondence to: loh@jilau1.colorado.edu (H.L.);
ye@jila.colorado.edu (J.Y.); cornell@jila.colorado.edu (E.A.C.)

[‡]Current address: Department of Chemistry and Chemical Biology,
Harvard University, Cambridge, MA 02138, USA

**Polar molecules are desirable systems for quantum simulations and cold chemistry. Molecular ions are easily trapped, but a bias electric field applied to polarize them tends to accelerate them out of the trap. We present a general solution to this issue by rotating the bias field slowly enough for the molecular polarization axis to follow but rapidly enough for the ions to stay trapped. We demonstrate Ramsey spectroscopy between Stark-Zeeman sublevels in $^{180}$Hf$^{19}$F$^+$ with a coherence time of 100 ms. Frequency shifts arising from well-controlled topological (Berry) phases are used to determine magnetic g-factors. The rotating-bias-field technique may enable using trapped polar molecules for precision measurement and quantum information science, including the search for an electron electric dipole moment.**

Quantum control of the rich internal structure of molecules may lead to advances in both fundamental and applied physics beyond those already gleaned from atoms (*1*). The same complex molecular structure makes it difficult to trap neutral molecules; in contrast, molecular ions can be easily trapped by time-dependent electric fields (*2-9*). However, the use of an ion trap has precluded taking advantage of the electric dipole moment of molecular ions, a vital ingredient of many experiments ranging from precision tests of fundamental physics (*2, 6, 10*) to quantum information (*11*).

The ability to polarize the molecules is, for example, a prerequisite for a search for the electron electric dipole moment (eEDM) with molecules (*12*): this search serves as a direct tabletop test of time-reversal violation (*13*) and as a probe of physics beyond the Standard Model (*14*). Trapped HfF$^+$ or ThF$^+$ molecular ions in the $^3\Delta_1$ state are excellent candidates for an eEDM search with several key advantages: small $\Omega$-doublet splitting for a built-in rejection of many systematic errors (*15-17*), a small magnetic moment for reduced magnetic field sensitivity (*18*), long coherence times for spectroscopy, and high internal electric fields for enhanced eEDM sensitivity (*2, 18-20*). In order to access the high internal field, the molecules need to be polarized in the laboratory frame by an applied bias electric field; however, their position in the ion trap ensures that the time-averaged electric field is zero.

To address this issue, we apply a rotating bias electric field, realizing the proposal of Ref. (*2*). The frequency of rotation $\omega_{\text{rot}}$ is low enough for the molecular polarization axis to adiabatically follow the electric field but high enough for the ions to stay trapped.

At the heart of our experiment is a linear Paul trap with six radial confinement electrodes (Fig. 1A-B). To form the rotating bias electric field $E_{\text{rot}}$, we apply a set of sinusoidal voltages to the radial electrodes, such that the phases on successive electrodes are each advanced by 60$^\text{o}$. $E_{\text{rot}}$ causes the molecular ions to rotate in a circular micromotion (Fig. 1B). Compared to all other forms of ion motion, including radiofrequency micromotion, the circular micromotion occurs on the fastest time scale. The fast time scale, combined with the spatial uniformity and large amplitude of $E_{\text{rot}}$, ensures that all the ions synchronously undergo the same circular micromotion. In this paper, the quantization axis is given by the instantaneous electric field.

For a precision measurement of the eEDM, it is advantageous to apply a magnetic field aligned along or against the bias electric field because the Zeeman interaction between the electron magnetic dipole moment and the magnetic field shifts the eEDM signal away from possible low-frequency noise sources. In our experiment, such a magnetic field can be implemented by combining a static magnetic-field gradient with the ions' circular micromotion (Fig. 1B). The magnetic field experienced by an ion over a rotation cycle can be decomposed into a spatially-dependent but roughly time-invariant component $\boldsymbol{B}_{\text{static}}(\boldsymbol{R})$, and a time-varying, almost spatially-independent component $\boldsymbol{B}_{\text{rot}}(t) = \boldsymbol{r}_{\text{rot}}(t)\, \partial B_\rho/\partial \rho$, where $\boldsymbol{r}_{\text{rot}}(t)$ is the circular micromotion and $B_\rho$ is the radial component of the magnetic field. $\boldsymbol{B}_{\text{static}}$ can be safely neglected, because the rotating bias electric field results in sensitivity to only $\boldsymbol{B}_{\text{rot}}$ (2).

In the presence of a bias electric field and a co-parallel magnetic field, the rotating-frame energy levels of the metastable $^3\Delta_1$ ($v = 0$, $J = 1$, $F = 3/2$) state of HfF$^+$ are split (Fig. 1C). The interaction between the molecules' electric dipole moment and the bias field gives rise to four pairs of Stark-split levels. The Stark spectrum in Fig. 1C is recorded by performing two-photon Raman transfer to populate ions in the desired $^3\Delta_1$ state from the ground $^1\Sigma^+$ state via an intermediate $^3\Pi_{0+}$ state. A particular Stark level-pair can thus be isolated by tuning the Raman transfer laser frequencies to the appropriate two-photon resonance.

For the eEDM measurement, we focus on either the uppermost ($u$) or lowermost ($l$) Stark pairs. In the absence of $B_{\text{rot}}$, the coupling from the electric field rotation results in an avoided crossing of the two eigenstates $\{|+\rangle, |-\rangle\}$ separated by the energy splitting $\Delta$. The eigenstates are equal superpositions of the spin projections, i.e. $|\pm\rangle_{B_{\text{rot}}=0} = \left(|m_F = 3/2\rangle \pm |m_F = -3/2\rangle\right)/\sqrt{2}$, where $m_F$ is the spin-projection quantum number. As the Zeeman energy increases relative to $\Delta$, the eigenstates of the Hamiltonian evolve from different superpositions of $|m_F = \pm 3/2\rangle$ to pure spin states (Fig. 1D). For sufficiently large magnetic fields, the energy eigenstates at high $B_{\text{rot}}$ are $|\pm\rangle = |m_F = \pm 3/2\rangle$ and their energies asymptote towards a linear Zeeman shift. In the presence of the avoided crossing, the rotating magnetic field is not merely an advantage for precision measurement; it is a necessity for experiments aiming to manipulate states of distinct $m_F$.

To probe the physics of the rotating-frame Hamiltonian within a single Stark pair, we perform Ramsey spectroscopy on the individual sublevels as a function of the rotating magnetic field. The Ramsey sequence always begins with populating ions in a single Stark pair, and then optically pumping away ions in the $|m_F = -3/2\rangle$ state using a $\sigma^+$-polarized depletion laser. For small $B_{rot}$ (i.e. small Zeeman splitting compared to $\Delta$), the depletion process acts like a $\pi/2$ pulse, leaving the ions in a superposition of the eigenstates $|\pm\rangle$ (Fig. 2A). At high $B_{rot}$, the $\pi/2$ pulse is instead executed by fast ramps of the rotating electric field to and from a lower magnitude (*21*). The full Ramsey sequence (Fig. 2B) is the same as that used for an eEDM measurement.

The number of ions in either one of the two $|m_F\rangle$ states after the second $\pi/2$ pulse is determined by photodissociating the remaining ions in the $^3\Delta_1$, $J = 1$ state, and counting the dissociated ions via a mass-resolved ejection sequence. The fractional population difference between the two $|m_F\rangle$ states oscillates as a function of the wait time $T$ between the two $\pi/2$ pulses at a frequency that is the avoided-crossing splitting (Fig. 2C). The decay time constant of the Ramsey fringe shows the ions maintaining coherence over a long period of 100(30) ms. The correspondingly high spectral resolution on the eEDM transition reduces vulnerability to potential systematic effects. The coherence time can potentially reach beyond 1 s, which is set by the $^3\Delta_1$ spontaneous-decay lifetime. The present limit most likely comes from ion-ion interactions.

Figure 3 shows the full avoided-crossing splittings for the upper and lower Stark pairs as a function of $B_{rot}$ with a rotating bias field of 11.6 V/cm. A fit to the data yields the magnitude of the magnetic g-factor, which is spectroscopically relevant for the eEDM experiment. We measured $g_F \equiv (g_F^u + g_F^l)/2 = +0.00306(10)$ and the difference in g-factors to be $\delta g_F = g_F^u - g_F^l = 0.00001(2)$. As expected, $g_F \ll 1$ due to the cancellation of orbital and spin angular momenta in the $^3\Delta_1$ state (*2*). Further, the magnetic g-factor is similar for both the upper and lower Stark manifolds, which is advantageous for suppressing systematic errors in an eEDM experiment. Note that a nonzero electron EDM would lead to a horizontal offset of the upper and lower curves in Fig. 3 from each other. The data shown represent our first eEDM measurement and already constrain its magnitude to be less than $1.5 \times 10^{-25}$ e·cm. A dedicated eEDM effort may lead to a statistical uncertainty of $1 \times 10^{-28}$ e·cm in a day (*21*).

A Ramsey-spectroscopy measurement of the electron EDM will determine its sign relative to the sign of the molecule's magnetic g-factor. Our tool for determining the sign of $g_F$ is the Berry phase (*22*).

The Berry phase is a dynamical phase shift imprinted on a quantum mechanical state due to the Hamiltonian having explicit time dependence. Under the influence of a rotating bias field, the ions can accumulate a finite Berry phase if the quantization axis is tilted away from the plane of rotation (*23*). In our technique the resultant Berry phase is much better controlled across the ion cloud compared to that for a trapped sample of neutrals (*24*, *25*).

To demonstrate this control, we explicitly incorporate its spectroscopic effect by giving the ions an impulse axial kick before performing the Ramsey sequence in a finite positive rotating magnetic field. The ions subsequently oscillate at the axial trap frequency. The Ramsey sequence is completed while the ions have only gone through half an axial oscillation, such that they are only present in either the z > 0 (z < 0) region for a upward (downward) kick. The downward-pointing (upward-pointing) trap field causes the quantization axis to tilt out of the plane of rotation, such that the solid angle $\Omega_{SA}$ subtended by the quantization axis is larger (smaller) than that without the kick (Fig. 4A). The resulting Berry phase $\phi_{Berry}$ manifests as a Berry energy $U_{Berry} = -\dfrac{\hbar \phi_{Berry}}{(2\pi)/\omega_{rot}} = \hbar m_F \dfrac{\omega_{rot}}{2\pi} \Omega_{SA}$ over one rotation cycle. For $g_F B_{rot} > 0$, the $m_F = 3/2$ level lies below the $m_F = -3/2$ level, so $U_{Berry}$ causes the two $m_F$ levels to move closer (split further apart) for an upwards (downwards) kick (see Fig. 4B). For a smaller (larger) energy splitting, the Ramsey fringe frequency is slower (faster), which looks like the Ramsey fringe has accumulated a positive (negative) initial phase shift. The two Ramsey fringes separately recorded with the ions kicked upwards (dots) and downwards (crosses) from the plane of rotation have a positive and negative phase shift respectively (shaded region of Fig. 4C), which means that $g_F B_{rot} > 0$. Because $B_{rot} > 0$, we determine the magnetic g-factor to be positive.

We have demonstrated coherent spectroscopy with molecular ions that are simultaneously polarized and trapped in a linear Paul trap with a rotating bias field. The rotating bias field and co-rotating magnetic field open up access to the manipulation of Stark-Zeeman sublevels within a hyperfine-rovibronic state. The long coherence time of a qubit encoded by Stark-Zeeman states, whose relative energies are much more stable (compared to rotational or even hyperfine levels) against Stark-shift-induced decoherence mechanisms, is useful for quantum information processing (*11*). For quantum simulation experiments, the direct ability to manipulate dipolar interactions with molecular ions potentially eliminates the need for using optical spin-dependent forces to achieve spin-spin couplings, as is the case for trapped atomic ions (*26-28*). Further, the 10-Hz-level high-resolution spectroscopy is directly relevant to precision tests of ab initio theory (*6*), time variations in fundamental constants (*29*) and symmetry violations (*30*).

**Acknowledgments:**

This work is supported by NIST, the Marsico Foundation and NSF grant #1125844. H. L. is partly funded by A*STAR (Singapore). K.-K. N. acknowledges a NIST/NRC Postdoctoral Fellowship. We thank Trent Fridey for his contributions and Daniel Gresh for discussions. We also recently became aware of a new eEDM result from the ACME collaboration (*36*).

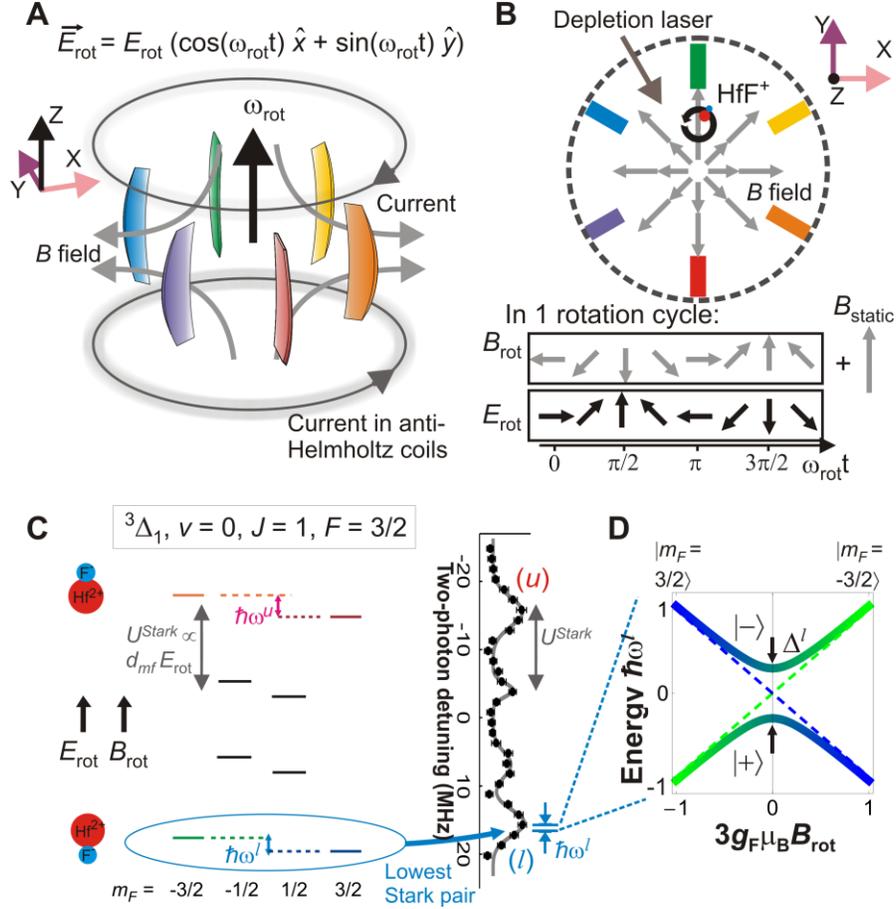

**Fig. 1. Rotating-bias-field technique.** (**A**) Three-dimensional view of a linear Paul trap (axial-confinement electrodes not shown). The colored radially-confining electrodes carry an additional set of phase-shifted sinusoidal voltages to create a rotating bias field $E_{rot}$. The pair of anti-Helmholtz coils creates a static magnetic-field gradient. (**B**) Top view of ion trap, with a depletion laser (see text) propagating between two radial electrodes. In one rotation cycle, the magnetic field (gray arrows) sampled by a molecular ion can be decomposed into a rotating component $B_{rot}$ and a time-invariant component $B_{static}$. (**C**) Rotating-frame HfF$^+$ energy level diagram for the $^3\Delta_1$ ($v = 0$, $J = 1$, $F = 3/2$) state in the presence of $E_{rot}$ and $B_{rot}$. $E_{rot}$ splits the levels into four spectroscopically isolated Stark pairs. (**D**) Energies of sublevels within a single Stark pair. The rotation of the electric bias field couples the two sublevels, turning the linear Zeeman splitting (dashed lines) into an avoided crossing (solid lines) split by $\Delta$ at the point of zero $B_{rot}$. For a typical measurement, the Stark splitting, rotation rate, Zeeman splitting, and avoided-crossing mixing are given by 10 MHz, 250 kHz, 100 Hz, and 30 Hz, respectively.

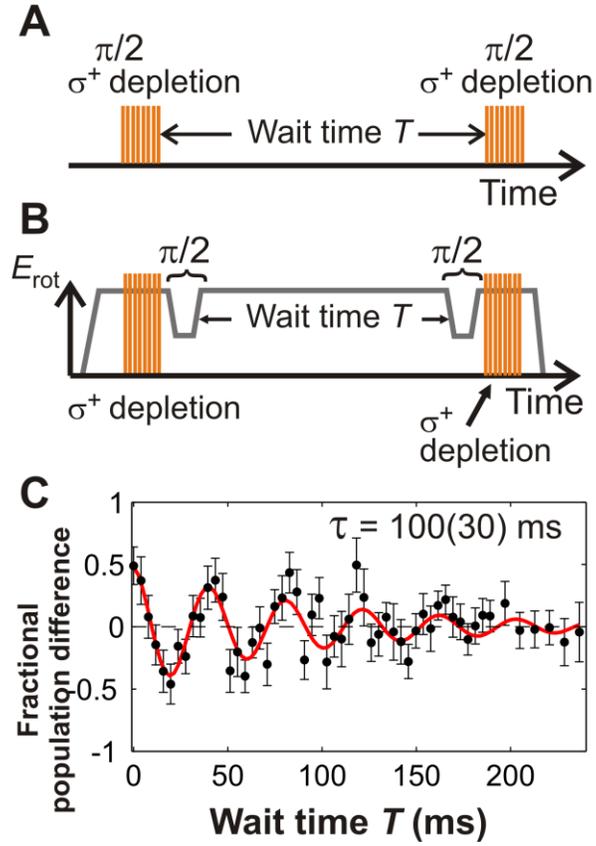

**Fig. 2. Ramsey spectroscopy in rotating fields.** Ramsey sequences (A) for $B_{rot} = 0$, where the $\sigma^+$-polarized depletion laser acts as both a polarizer and a $\pi/2$ pulse; (B) for high $B_{rot}$, where ramps in $E_{rot}$ act as $\pi/2$ pulses. (C) Fractional difference of ions in $|m_F = 3/2\rangle$ versus that in $|m_F = -3/2\rangle$ as a function of the Ramsey wait time $T$, taken at $B_{rot} = 0$. The data is fit to a sinusoidal function with an exponential-decay time constant $\tau$.

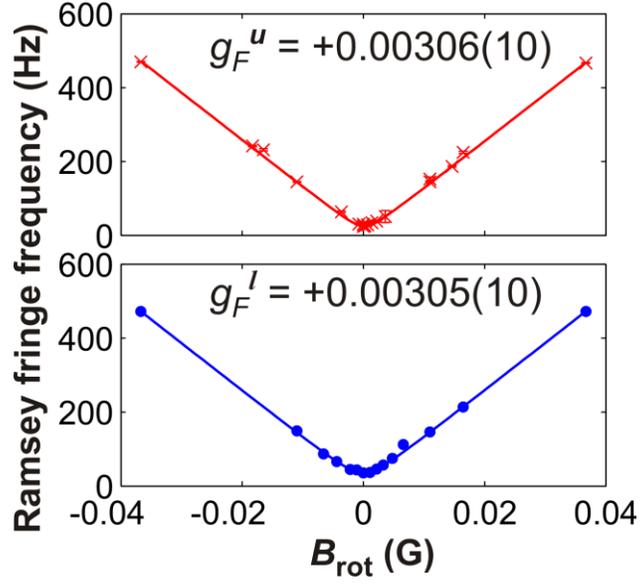

**Fig. 3. Measurement of the avoided-crossing splitting.** Avoided crossing for $E_{\text{rot}} = 11.6$ V/cm, experimentally mapped out using Ramsey spectroscopy for the upper (crosses) and lower (dots) Stark pairs. The upper and lower magnetic g-factors $g_F$ that have been obtained from the fits (solid lines) are nearly identical and quite small.

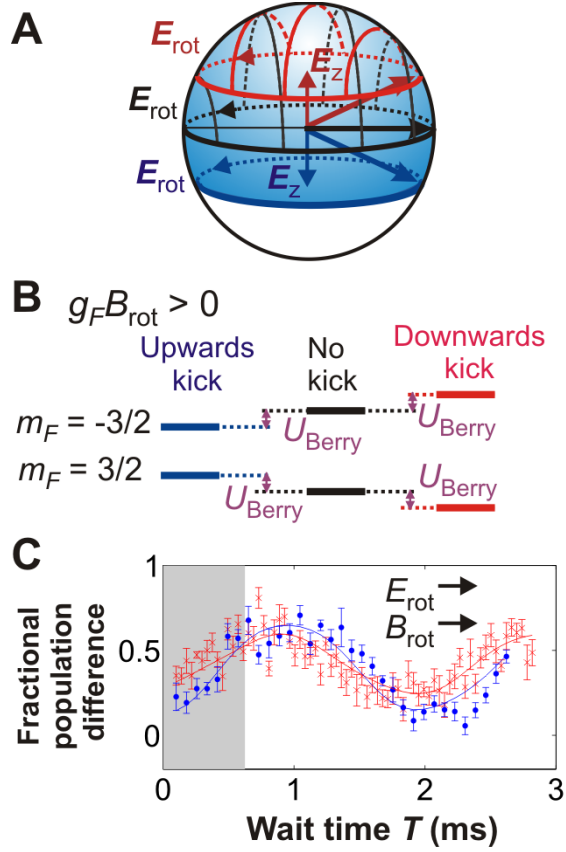

**Fig. 4. Determination of the sign of the magnetic g-factor using Berry phase.** (**A**) Ions rotating in the $z = 0$ plane subtend a solid angle of $2\pi$ (black hatched hemisphere). As the ions are kicked above (below) the plane of rotation, they sample a force-restoring axial electric field $E_z$. The quantization axis thus tilts away from the plane of rotation, subtending a larger (smaller) solid angle indicated by the blue shaded (red hatched) region. (**B**) For $g_F B_{\text{rot}} > 0$, the additional Berry energy $U_{\text{Berry}}$ from the axial kicks causes the $|m_F = \pm 3/2\rangle$ levels to move closer together or farther apart in energy. (**C**) Ramsey fringes recorded for $B_{\text{rot}} > 0$ (i.e. $\boldsymbol{B}_{\text{rot}}$ in phase with $\boldsymbol{E}_{\text{rot}}$), where the ions have been kicked above (blue dots) or below (red crosses) the plane of rotation. The relative phase shifts mean $g_F > 0$. The gray shaded region corresponds to half an axial trap oscillation, which is the time over which the ions are exclusively in either the $z > 0$ or $z < 0$ region. The solid lines are fits to the data, incorporating the frequency chirp expected as the ions oscillate axially.

# Supplementary Material

## Materials and Methods

The experiment sequence starts with the creation of neutral HfF molecules from ablating a Hf rod in the presence of a gas comprising 1% $SF_6$ and 99% Ar. The gas is released into the vacuum chamber through the opening of a pulsed valve (800 μm orifice, 830 kPa backing pressure) for 180 μs. The neutral HfF molecules are cooled by supersonic expansion to a translational and rotational temperature of ~10 K. To create $HfF^+$, we cross two pulsed lasers (309 nm, 368 nm) with the supersonic beam downstream of the ablation to excite ground state neutral HfF molecules to a Rydberg state with an excited ion core. The Rydberg molecules can then autoionize to form $^{180}Hf^{19}F^+$ ions, 35% of which are in the $^1\Sigma_0^+$ ($v'' = 0$, $J'' = 0$) ground rovibronic state (*31*).

The ionization process takes place inside the center of the ion trap, so that upon ionization, we can immediately stop the ions with a pulsed stopping field and then trap them with the linear quadrupole Paul trap. The rotating bias field is ramped on by applying an additional set of sinusoidal voltages to the radial-confinement electrodes on top of the trapping radiofrequency (RF) and direct current (DC) voltages. Under the influence of the rotating bias field and trapping fields, the ions undergo a combination of secular harmonic motion at $\omega_{sec}/(2\pi) = 5$ kHz, trap-RF-induced micromotion at $\omega_{RF}/(2\pi) = 50$ kHz and rotating-bias-field-induced circular micromotion at $\omega_{rot}/(2\pi) = 253$ kHz in the radial plane. (In the axial direction, the ions experience only secular harmonic motion at $\omega_z/(2\pi) = 0.8\text{-}3.6$ kHz.)

In the presence of the rotating bias field, we populate ions in a desired Stark level-pair of the $^3\Delta_1$ ($v = 0$, $J = 1$, $F = 3/2$) state using two Raman transfer lasers. The first transfer laser at 900 nm drives the $^3\Pi_{0+}$ ($v' = 1$, $J' = 1$) ← $^1\Sigma_0^+$ ($v'' = 0$, $J'' = 0$) transition while the second transfer laser at 986 nm drives the $^3\Pi_{0+}$ ($v' = 1$, $J' = 1$) → $^3\Delta_1$ ($v = 0$, $J = 1$, $F = 3/2$) transition (*32*). Both lasers are detuned by 160 MHz from their respective one-photon resonances to avoid populating the $^3\Pi_{0+}$ excited state. Light from both lasers is simultaneously directed onto the ions through the top electrode of the ion trap for 1.2 ms, during which the trapped ions' axial motion Doppler-shift the transfer laser frequencies to sweep through the two-photon detuning. The desired Stark level-pair is populated when the two-photon resonance condition is fulfilled.

To deplete ions from the $|m_F = -3/2\rangle$ sublevel within a Stark level-pair, we direct a $\sigma^+$-polarized depletion laser, tuned to resonance with the $^3\Pi_{0^+}$ ($v' = 1$, $J' = 1$) $\rightarrow$ $^3\Delta_1$ ($v = 0$, $J = 1$, $F = 3/2$) transition, between two radially-confining trap electrodes onto the ions. The depletion laser is stroboscopically turned on, with a duty cycle of 20%, whenever the rotating quantization axis lies roughly parallel to the direction of laser propagation. 250 rotation cycles (about 1 ms) are needed to fully deplete the ions from one of the two $m_F$ sublevels. In the absence of a rotating magnetic field, the depletion laser acts as a $\pi/2$ pulse because the remaining ions are left in a single $|m_F\rangle$ sublevel, which is an equal superposition of the eigenstates $|\pm\rangle_{B_{rot}=0}$.

For experiments where a rotating magnetic field is needed, the static magnetic field gradient is applied prior to ionization by running current through a pair of anti-Helmholtz coils positioned just outside the ion trap. The radial magnetic field gradient can be increased up to 1.65 G/cm, which corresponds to an effective rotating magnetic field up to 0.037 G for a circular micromotion radius of 0.22 mm obtained with a rotating electric field of 11.6 V/cm. Reversing the sign of the rotating magnetic field from positive to negative is accomplished by changing the direction of applied current, such that the magnetic field gradient points radially outwards instead of inwards. The static magnetic field component $\boldsymbol{B}_{\text{static}}$ acting on our rotating eigenstates generates nothing but a negligible ($< 10^{-3}$) phase modulation over a period of rotation. A displacement of the null of the static magnetic-field quadrupole relative to the trap center along the radial or axial directions therefore also exerts a negligible effect because it merely affects $\boldsymbol{B}_{\text{static}}$ and not $\boldsymbol{B}_{\text{rot}}$, and because the quantization axis is nearly entirely defined by electric and not magnetic fields. Ref. (*2*) offers a more in-depth analysis of the negligible effects of static magnetic-field gradient misalignments on precision spectroscopy.

Finally, to detect the ions in a particular $m_F$ sublevel after the Ramsey sequence, we use the depletion strobe-sequence to optically pump away ion population in the undesired $m_F$ sublevel without depopulating the desired sublevel, before photodissociating the ions remaining in the $J = 1$ state with 286 nm and 266 nm pulsed lasers. The dissociated ions are counted on a microchannel plate as $Hf^+$ atomic ions, mass-resolved from $HfF^+$ in time-of-flight. The 286 nm photon drives a bound-bound transition within the $HfF^+$ molecule and provides rotational-state selectivity to the dissociation. A manuscript on the photodissociation process is being prepared.

## Supplementary Text

Rotating-frame Hamiltonian

Within either the uppermost or lowermost Stark level-pair, the two sublevels $\{|m_F = 3/2\rangle, |m_F = -3/2\rangle\}$ could be coherently coupled to each other via a three-photon transition. Alternative to using three radiofrequency photons, we can take advantage of the coupling (via third-order perturbative process) already provided by the rotation of the electric field.

The rotating-frame Hamiltonian is expressed in the $\{|m_F = +3/2\rangle, |m_F = -3/2\rangle\}$ basis as follows (*2*):

$$H = \begin{pmatrix} -\frac{3}{2} g_F \mu_B B_{rot} & \frac{\Delta}{2} \\ \frac{\Delta}{2} & \frac{3}{2} g_F \mu_B B_{rot} \end{pmatrix}, \quad (S1)$$

where $g_F$ is the magnetic g-factor for the $F = 3/2$ state, $\mu_B$ is the Bohr magneton, $B_{rot}$ is the rotating magnetic field and $\Delta$ is the splitting between the energy eigenstates at $B_{rot} = 0$. The coupling energy $\Delta$, which arises from the rotation of the quantization axis, is very similar for both the uppermost and lowermost Stark level-pairs $\Delta^{u/l}$. For a three-photon transition necessary to connect the $m_F = \pm 3/2$ states, the average of $\Delta$ for the two Stark pairs is proportional to the ratio of the rotation frequency $\omega_{rot}$ to the Stark energy $d_{mf}E_{rot}$, raised to the power of three:

$$\Delta = \frac{1}{2}(\Delta^l + \Delta^u) = 27\hbar\omega_{ef}\left(\frac{\hbar\omega_{rot}}{d_{mf}E_{rot}}\right)^3. \quad (S2)$$

$\varDelta$ is also proportional to the energy splitting between the parity eigenstates at zero electric field $\omega_{ef}$, which sets the scale of the coupling between states with opposite sign $\Omega$ (where $\Omega$ is the projection of total angular momentum onto the molecular axis).

There is a slight frequency difference between $\varDelta$ for the two Stark pairs, which comes from interactions with the nearest $F = 1/2$ hyperfine level, detuned from the $F = 3/2$ manifold by $E_{HF}$. Specifically, the upper Stark pair is more repelled from the $F = 1/2$ levels than the lower Stark pair. Perturbation theory on the Hilbert space of twelve $|F, m_F, \Omega\rangle$ sublevels in the $^3\Delta_1$ ($v = 0$, $J = 1$) rotational state yields the following expression for $\delta\varDelta$:

$$\delta\Delta = \frac{1}{2}\left(\Delta^l - \Delta^u\right) = \frac{81}{8}\hbar\omega_{ef}\left(\frac{\hbar\omega_{rot}}{d_{mf}E_{rot}}\right)^3\left(\frac{d_{mf}E_{rot}}{E_{HF}}\right). \tag{S3}$$

We note that the numerical coefficients in Eqs. (S2)-(S3) were calculated numerically for Ref. (*2*), but there was an error in transcribing the results to that manuscript. The expressions presented here are analytical results from perturbation theory and are consistent with revised numerical calculations.

Figure S1 shows an experimental verification of the third-order dependence of $\varDelta$ on $1/E_{rot}$, measured by repeating the Ramsey spectroscopy technique for different values of $E_{rot}$. The lines come from a simultaneous fit to $\varDelta^u$ and $\varDelta^l$ using Eqs. (S2)-(S3). The sole adjustable parameter, $\omega_{ef}/(2\pi)$, is determined to be 830(50) kHz. $\omega_{ef}/(2\pi)$ was independently determined in Ref. (*32*) to be 740(20) kHz, and the two measurements agree to within two standard deviations.

We take advantage of the strong dependence of the avoided-crossing splitting and its eigenstates on $E_{rot}$ (Fig. S2) to execute $\pi/2$ pulses at high magnetic fields. For a given finite magnetic field, we can ramp $E_{rot}$ to an appropriate lower value and back in order to project the initial $|m_F\rangle$ state onto an equal superposition of the energy eigenstates $|\pm\rangle$.

The full avoided-crossing energy measured as a function of $B_{rot}$ for either the upper or lower Stark pair, as shown in Fig. 3, is fit to the following function:

$$\frac{\omega^{u/l}}{2\pi} = 2\sqrt{\left(\frac{\Delta^{u/l}}{2h}\right)^2 + \left(\frac{3}{2}\mu_B g_F^{u/l}\left(B_{rot} - B_{offset}^{u/l}\right)\right)^2} \, , \qquad (S4)$$

where $B^{u/l}{}_{offset}$ is a magnetic field offset that shifts the avoided crossing away from zero along the horizontal axis. The fit parameters yield $\Delta^u/h$ = 25(1) Hz and $\Delta^l/h$ = 36(1) Hz, which are in good agreement with that expected from Eqs. (S2)-(S3). The magnetic field offsets fit to $B^u{}_{offset}$ = 0.15(11) mG and $B^l{}_{offset}$ = 0.02(9) mG. The average of both offsets is roughly that expected from driving currents up and down the radial confinement electrodes to provide the rotating bias field. The difference between the two offsets, which is proportional to the eEDM signal, is consistent with zero. Combined with the theoretical value for the effective electric field on the electron $E_{eff}$ = 24 GV/cm (*19, 33*), the difference in $B_{offset}$ yields a preliminary eEDM limit of 1.5 x 10$^{-25}$ e·cm.

The eEDM statistical uncertainty can potentially average to 1 x 10$^{-28}$ e·cm in 12 hours assuming 4 detected ions making the Ramsey transition, a 150 ms coherence time, and a 4 Hz repetition rate for the experimental cycle. The systematic uncertainties have been estimated in Ref. (*2*) to be on the order of 10$^{-29}$ e·cm.

Sign of the magnetic g-factor

When the rotating quantization axis is tilted away from the plane of rotation by a small angle $\alpha$, such that one rotation traces out a cone with opening half-angle $\pi/2 - \alpha$ and solid angle $\Omega_{SA}$, the ions in a given $m_F$ level accumulate a Berry phase shift (*22*)

$$\phi_{Berry} = -m_F \Omega_{SA} = -m_F 2\pi(1 - \sin(\alpha)) \, . \qquad (S5)$$

The resulting phase shift on the spin state $|m_F\rangle$, over one rotation period $\tau = (2\pi)/\omega_{rot}$, is correspondingly $e^{-i\omega_m \tau} e^{i\phi_{Berry}} = e^{-i(\omega_m - \frac{\phi_{Berry}}{\tau})\tau}$, where $\omega_m$ is the Zeeman frequency. The Berry phase, modulo an integer multiple of $\pi$, translates to a frequency shift of (*2*)

$$\omega_{Berry} = -\frac{\phi_{Berry}}{\tau} = m_F \frac{\omega_{rot}}{2\pi}(2\pi)(-\sin(\alpha)) \approx -m_F \omega_{rot} \alpha. \tag{S6}$$

In the case of the quantization axis rotating about a small tipping angle ($\pi/2-\alpha \ll 1$), one can describe the Berry-phase frequency shift instead as a sort of low-frequency AC Stark perturbation (*34, 35*). For the maximal tipping angle of $\pi/2$, however, that perturbative approach does not yield useful intuition. To verify the simple Berry phase result (S6) above, one may instead laboriously rediagonalize the problem in a rotating frame (*23*).

To impart a Berry phase, we deliberately kick the ions above or below the plane of rotation. The maximum tilt angle of the quantization axis is then given by $\alpha = E_{z,max}/E_{rot}$, where $E_{z,max}$ is the maximum axial trapping field experienced by the ions as they oscillate away from the plane of rotation (i.e. $E_{z,max} < 0$ for an upward kick). The diagonal terms of the Hamiltonian in Eq. (S1) in turn become $\mp m_F g_F \mu_B B_{rot} + \omega_{Berry} \sin(\omega_z t)$, where $\sin(\omega_z t) > 0$ in half an axial trap cycle.

Relating the energies of the sublevels back to experimental parameters, we obtain the following dependence for the sign of $\omega_{Berry}$:

$$\text{sign}(\omega_{Berry}) = + \text{sign}(m_F)\,\text{sign}(\omega_{rot})\,\text{sign(kick)}, \tag{S7}$$

where $\omega_{rot}$ is positive (negative) if the quantization axis rotates counter-clockwise (clockwise), and sign(kick) is positive (negative) if the ions are kicked upwards (downwards).

Our convention for the sign of the magnetic g-factor is defined in Eq. (S1). Therefore, the sign of the Zeeman energy is

$$\text{sign(Zeeman energy)} = -\text{sign}(m_F)\,\text{sign}(g_F)\,\text{sign}(|\mu_B|)\,\text{sign}(B'), \tag{S8}$$

where $B'$ parameterizes the static magnetic field as $\boldsymbol{B} = B'z\hat{z} - \frac{B'}{2}\rho\hat{\rho}$. Since $\boldsymbol{B}_{rot} = \boldsymbol{r}_{rot}\,\partial B_\rho/\partial \rho$ and $\boldsymbol{r}_{rot} \propto -\boldsymbol{E}_{rot}$ for circular motion, $\text{sign}(B') = \text{sign}(B_{rot})$.

It then follows from Eqs. (S7) and (S8) that

$$\text{sign}(g_F) = -\text{sign}(B')\,\text{sign}(\omega_{rot})\,\text{sign(fringe frequency)}\,\text{sign(kick)}, \tag{S9}$$

where sign(fringe frequency) is positive if the fringe frequency increases. Applying Eq. (S9) to the data presented in Fig. 4, taken with $B' > 0$ and $\omega_{\text{rot}} > 0$, $g_F$ is determined to be positive.

Having understood the effect of Berry phase, we note that during an actual precision measurement, care must be taken to collect data asynchronously with any unintended initial axial slosh of the ions' center of mass, so as to minimize systematic errors associated with Berry phase. This should not pose a major technical challenge.

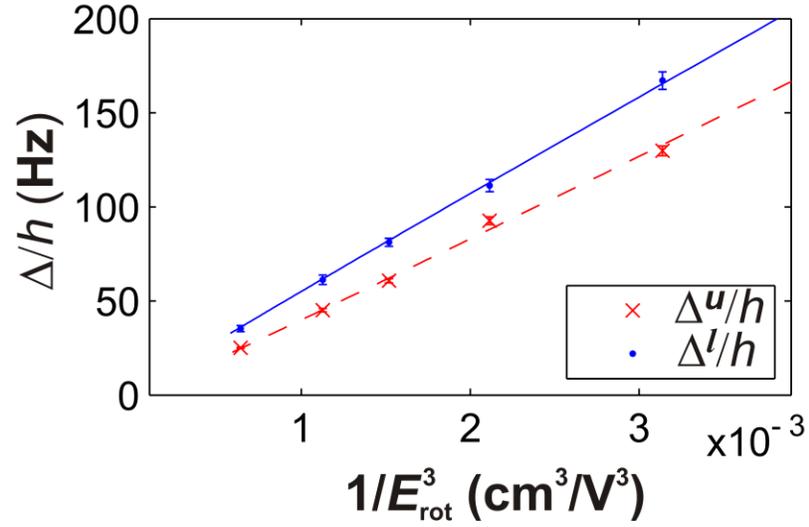

**Fig. S1. Measurement of avoided-crossing splitting at zero rotating magnetic field.**

Ramsey fringe frequency ($\Delta/h$) measured as a function of the magnitude of the rotating-bias-field $E_{rot}$ (red crosses for the upper Stark pair and blue dots for the lower Stark pair). The lines are a single-adjustable-parameter fit to the data using theory describing the rotating-frame Hamiltonian (*2*).

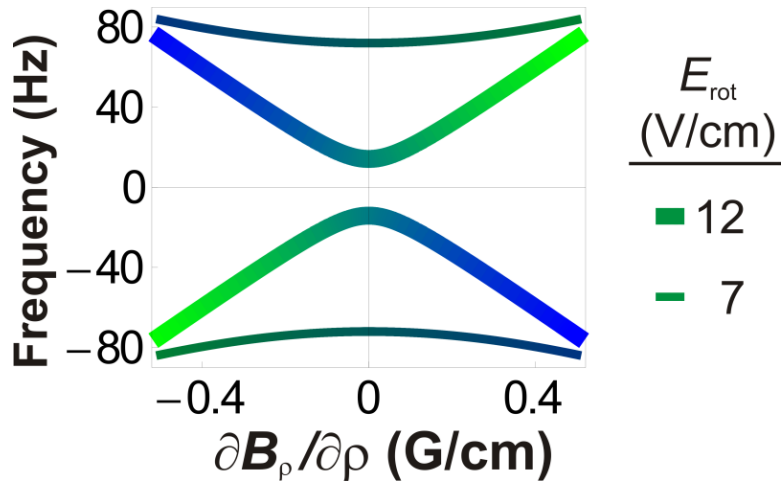

**Fig. S2. Avoided-crossing splitting as a function of applied magnetic-field gradient.**
Theory curves describing the avoided crossing, plotted with two different line thicknesses for the two values of $E_{rot}$. The eigenstates are depicted as superpositions of the colors blue (for $|m_F = 3/2\rangle$) and green (for $|m_F = -3/2\rangle$). For a given magnetic-field gradient (e.g. 0.4 G/cm), the eigenstates can change from being almost pure $|m_F\rangle$ states at $E_{rot} = 12$ V/cm to near-equal superpositions of the two $|m_F\rangle$ states at $E_{rot} = 7$ V/cm.